\documentclass[12pt,preprint]{aastex}
\usepackage{color}

\shorttitle{superflare SEP}
\shortauthors{Takahashi et al.}

\begin{document}
\title{Scaling Relations in Coronal Mass Ejections and Energetic Proton Events associated with Solar Superflares}

\author{
Takuya Takahashi\altaffilmark{1},
Yoshiyuki Mizuno\altaffilmark{2},
Kazunari Shibata\altaffilmark{1}}
\email{takahasi@kusastro.kyoto-u.ac.jp}

\altaffiltext{1}{
Kwasan and Hida Observatories, Kyoto University,
Yamashina, Kyoto 607-8471, Japan.}
\altaffiltext{2}{
Faculty for the Study of Contemporary Society, Kyoto Women's University, 
35 Kitahiyoshi-cho, Imakumano, Higashiyama-ku, Kyoto 605-8501 Japan}

\begin{abstract}
In order to discuss the potential impact of solar 'superflares' on space weather, we investigated statistical relations among energetic proton peak flux with energy higher than $ 10 \rm MeV$ ($F_p$), CME speed
near the Sun ($V_{CME}$) obtained by {\it SOHO}/LASCO coronagraph and flare soft X-ray peak flux in 1-8\AA band ($F_{SXR}$) during 110 major solar proton events (SPEs) recorded from 1996 to 2014. The linear regression fit results in the scaling relations $V_{CME} \propto F_{SXR}^\alpha$, $F_p\propto F_{SXR}^\beta$ and $F_p\propto V_{CME}^\gamma$ with $\alpha = 0.30\pm 0.04$, $\beta = 1.19 \pm 0.08$ and $\gamma = 4.35 \pm 0.50$, respectively. On the basis of simple physical assumptions, on the other hand, we derive scaling relations expressing CME mass ($M_{CME}$), CME speed and energetic proton flux in terms of total flare energy ($E_{flare}$) as, $M_{CME}\propto E_{flare}^{2/3}$, $V_{CME}\propto E_{flare}^{1/6}$ and $F_{p}\propto E_{flare}^{5/6}\propto V_{CME}^5$, respectively. We then combine the derived scaling relations with observation, and estimated the upper limit of $V_{CME}$ and $F_p$ to be associated with possible solar superflares. 
\end{abstract}

\keywords{Sun: coronal mass ejections ---Sun: flares --- solar-terrestrial relations --- Stars: activity}

\section{Introduction}
Solar flares are the biggest explosion in the solar system where magnetic field energy stored in the active region corona is rapidly released through magnetic reconnection process \citep{shibm2011,hud2011}. During flares, coronal plasma is sometimes ejected out into the interplanetary space (coronal mass ejections; CMEs) \citep{JGR1986,gopa2009}. Significant portion of active region magnetic field energy released during flares is converted to the kinetic energy of CMEs \citep{emslie2012}. 
The resultant plasma and magnetic field structures detected at 1 AU are called interplanetary CMEs (ICMEs)\citep{zhang2007}. When helical magnetic field of ICME ejecta (magnetic cloud) or draped magnetic field in the interplanetary sheath ahead of magnetic cloud have strong southward component, geomagnetic storms occur \citep{JGR1982,JGR1988}.


Energetic protons are accelerated both at CME-driven shocks and flare site \citep{reames1996,RS2009}. In the case of shock acceleration mechanism, the efficiency of particle acceleration depends on shock Mach number and its normal direction, and particles are known to be most efficiently accelerated when the shocks are quasi-parallel \citep{JGR1984a,JGR1985}. Accelerated particles arrive at Earth when Earth is magnetically well connected to the shock front (solar proton event; SPE). Shock acceleration mechanism is generally thought to be dominant of the two, but extremely high energy protons (with energy $\sim$ GeV) accelerated at the flare site right after the flare onset are discussed to be responsible for the prompt component of Ground Level Enhancement (GLE) events \citep{asch2012}.

Large SPEs are often associated with large solar flares or fast CMEs \citep{gopa2004}.
The largest SPE after 1970 in terms of $E >$ 10 MeV proton peak flux occurred on 4 August 1972. The estimated $E >$ 10 MeV proton peak flux is higher than $\rm 6\times10^4$ pfu (particle flux unit; particles sr$^{-1}$ cm$^{-2}$ s$^{-1}$) \citep{kurt2004}. Modern extreme events that occurred on 23 July 2012 recorded the peak $E >$ 10 MeV proton flux of $6.5\times10^4$ pfu observed with STEREO-A space craft when interplanetary shock wave passed the space craft \citep{gopa2014}.

CMEs and SPEs are the two main drivers of hazardous space weather outcomes such as potential radiation hazards for space astronauts, geomagnetic storms and telecommunication failures \citep{AJS1861,JGR2003}. On the other hand, fast CMEs and intense energetic protons might have had important influence on ancient terrestrial environment through chemical processes in the terrestrial atmosphere \citep{air2016}. Young stars (like our ancient Sun) which rotate very rapidly are known to frequently produce superflares (10-$10^4$ times more energetic flares than the largest solar flares ever observed) possibly due to its active dynamo. Recent observation by Kepler satellite revealed that some solar-type stars with rotation period longer than 10 days can also produce superflares though not very frequent \citep{mae2012}.

The most widely used index of solar flare magnitude is soft X-ray (SXR) peak flux monitored by GOES satellite in 1-8\AA$~$passband. The proportionality between hard X-ray (HXR) fluence and SXR peak flux of indivisual flares is known as Neupert effect, and it is thought to be a casual index of released magnetic field energy during flares \citep{neu1968}. The solar flares are known to follow the frequency distribution of power-law form in terms of SXR peak flux \citep{yas2006}.
The largest ever solar flare observed in X-ray that occurred on 4 November 2003 saturated the {\it GOES} X-ray detector in 1-8\AA$~$passband. The estimated flare class by linear extrapolation is X28, while X-ray class estimation based on ionosphereic response resulted in X$45\pm5$ \citep{GRL2004}. Evidence of a spike of carbon-14 isotope between 774-775 is discovered from tree rings indicative of huge solar energetic particle event driven by solar superflares \citep{miya2012}. 
\citet{gopa2010a} estimated maximum flare energy to be of order $10^{35}$ erg (flare of $\sim$X1,000 class) based on observed maximum magnetic field strength in a sunspot and largest AR size. They estimated maximum CME speed associated with such a largest class of flares to be 7,200 km s$^{-1}$ assuming the CME mass to be of order $10^{18}$ g \citep{gopa2011}.

There are studies on statistical relations among flare SXR peak flux observed with {\it GOES} ($F_{SXR}$), CME speed near the Sun ($V_{CME}$) and $E>10$ MeV proton peak number flux ($F_p$) from various datasets. $F_p$ is known to be correlated both with $V_{CME}$ and $F_{SXR}$ \citep{gopa2011,gopa2003b}. From 19 SPEs occurred during maximum to minimum of solar cycle 23, \citet{gopa2003a} reported statistical relation, $F_p\propto V_{CME}^{3.7}$ and $F_p\propto F_{SXR}^{0.63}$, respectively. Also, the correlation coefficients of $F_{SXR}$ and $V_{CME}$ for events during 1996-2007 and GLE events are reported to be 0.37 and 0.50, respectively \citep{yas2009,gopa2012}. The correlation between $F_{SXR}$ and kinetic energy of CME ($K=M_{CME}V_{CME}^2/2$ with $M_{CME}$ being CME mass estimated from coronagraph observations) are also reported \citep{gopa2009}.

In this paper, we analyzed the statistical relations among CME speed, peak energetic proton flux and flare SXR peak flux from a single list of events, i.e. 110 SPEs recorded from 1996 to 2014 whose peak proton flux in the E $ >10 \rm MeV$ channel of GOES satellite exceeded 10 pfu. We derive scaling relations among CME speed, energetic proton flux and flare SXR peak flux on the basis of simple assumptions and compared with observation.

Finally, we estimate how fast CMEs will be, and how intense proton flux will come during possible solar superflare events based on the combination of the scaling relations and observation. 

\section{Dataset}
A total of 143 major SPEs were recorded from 1996 to 2014 whose peak proton flux in the E $ >10 \rm MeV$ channel of {\it GOES} satellite exceeded 10 pfu. The events are listed in CDAW major SEP event list page.\footnote{http://cdaw.gsfc.nasa.gov/CME\_list/sepe/} For 110 out of 143 SPEs, both flare SXR peak flux and CME speeds near the Sun accompanied with the SPEs are determined with X-ray Sensor on board the {\it GOES} satellite and Large Angle and Spectrometric Coronagraph (LASCO; Brueckner et al. 1995) on board the Solar and Heliospheric Observatory ({\it SOHO}; Domingo et al. 1995), respectively. For 80 out of these 110 SPEs, the mass of accompanied CMEs are also determined with LASCO. The estimated speed and mass of CMEs are obtained from CDAW Data Center CME catalog \citep{yas2004}, which is also available online.\footnote{http://cdaw.gsfc.nasa.gov/CME\_list/}

\section{Statistical relations among flare SXR peak flux, CME speed and energetic proton flux}
We study statistical relation among $F_{SXR}$ and $F_p$ during 110 major SPEs recorded between 1996 and 2014. Most CMEs, namely 92 out of 110, were 'halo' ones. A 'halo' CME is an expanding plasma of CME that appears to form a halo of enhanced brightness completely surrounding the occulting disk when observed with a coronagraph \citep{how1982}. The estimation of speed and mass of halo CMEs by coronagraph observations generally contains large uncertainty. We neglect the uncertainty of speed and mass estimation based on coronagraph observations throughout the analysis.

In Figure 1, we show correlation plot of $F_p$ and $F_{SXR}$. A regression line is drawn as a solid line to fit the log-log data plot. We used ordinary least squares (OLS) bisector method for linear regressions hereafter, which is suitable for the discussion of underlying functional relation between two quantities \citep{iso1990}. The correlation coefficient was $r = 0.41$, and the linear regression fit gives $F_p \propto F_{SXR}^\beta$ with $\beta = 1.19\pm 0.08$. 
5 out of 110 SPEs had $F_p$ larger than $10^4$ pfu, and 4 out of the five were associated with X class flares.

$F_p$ also correlates with $V_{CME}$ in our data set(Figure 2). We note here that throughout the paper, the CME speed $V_{CME}$ refers to the estimated speed of CME near the Sun based on observation with {\it SOHO}/LASCO. The average CME speed of all the 110 events was 1566 km~s$^{-1}$, and the average CME speed of 5 most intense SPEs was 2016 km~s$^{-1}$.
The correlation coefficient between $F_p$ and $V_{CME}$ for our dataset was $r = 0.45$, and the linear regression result gives $F_p \propto V_{CME}^\gamma$ with $\gamma = 4.35 \pm 0.50$. The event with $F_p =$ 1860 pfu and $V_{CME} = $882 km~s$^{-1}$ shown as unfilled circle in Figure 2 seems to be an outlier, and the analysis without the event makes the correlation coefficient a little bit higher and the slope a little steeper, namely, $r = 0.48$ and $\gamma=4.50\pm 0.48$, respectively. The event shown as unfilled circle was associated with an X7.1 class solar flare, and the CME was a Halo one.

Figure 3 shows correlation plot between $V_{CME}$ and $F_{SXR}$. The correlation coefficient was 0.42, and the linear regression result gives $V_{CME} \propto F_{SXR}^\alpha$ with $\alpha = 0.30\pm 0.04$. In 80 SPEs out of 110 analyzed here, CME mass $M_{CME}$ is also estimated\footnote{http://cdaw.gsfc.nasa.gov/CME\_list/}. Correlation between $M_{CME}$ and $F_{SXR}$ in the 80 SPEs was very poor with $r = -0.02$, possibly due to large uncertainty in mass estimation of halo CMEs.

\section{Scaling relations between flare magnitude, CME speed, CME mass and peak proton flux}
We try to express CME mass ($M_{CME}$), CME speed ($V_{CME}$) and energetic proton peak flux ($F_{p}$) as a power-law form of total released energy during flares ($E_{flare}$) based on three simple physical assumptions. First, we assume the CME mass is the sum of the mass within gravitationally stratified active region (AR) corona,
\begin{equation}
M_{CME}=L^2\int_0^L\rho_0 \exp(-\frac{z}{H})dz \sim \rho_0 L^2H
\end{equation}
where $\rho_0,L$ and $H$ are the density at the base of the AR corona, the length scale of flaring AR and the pressure scale hight, respectively. We implicitly assumed AR corona size $L$ is much larger than the coronal scale height $H$, which is suitable for large AR where large flares can occur.

Next, we assume CME kinetic energy is proportional to the total energy released during the flare, which is also a constant fraction $f$ of AR magnetic field energy \citep{emslie2012},
\begin{equation}
E_{CME}=\frac{1}{2}M_{CME}V_{CME}^2\propto E_{flare}=f\frac{1}{8\pi}B_0^2 L^3
\end{equation}
where $E_{flare}$ and $E_{CME}$ are the total released energy during flares and CME kinetic energy respectively. $B_0$ is the active region magnetic field strength. Typical magnetic field strength of sunspots is of the order of $~3000$ G.

The first and second assumptions (equations (1) and (2)) lead to the following relations.
\begin{equation}
M_{CME}\propto E_{flare}^{2/3},
\end{equation}
\begin{equation}
V_{CME}\propto E_{flare}^{1/6}.
\end{equation}

\citet{aar2011} studied CME/flare pairs observed with LASCO and {\it GOES} occurred from 1996 to 2006 and found the statistical relationship $M_{CME} \propto F_{SXR}^{0.7}$. \citet{aar2012} further discussed the statistical relation of CME mass $M_{CME}$ and energy released in the form of SXR during flares $E_{SXR}$ of the form $M_{CME} = K_M E_{SXR}^\delta$, where $K_M = (2.7\pm1.2)\times 10^{-3}$ in cgs units, and $\delta = 0.63\pm 0.04$. Such observations seem to be consistent with our scaling relation of $M_{CME}\propto E_{flare}^{2/3}$. Very interestingly, such scaling relation between $M_{CME}$ and $F_{SXR}$ obtained from solar flare statistics is consistent with mega-flare observation on young T Tauri star implying the scaling relation holds in a very wide energy range, that is more than 10 orders of magnitude in flare energy \citep{aar2012}.

We then try to relate energetic proton peak flux $F_p$ with flare energy $E_{flare}$. We assume that the total kinetic energy of solar energetic protons $E_p$ is proportional to flare energy, and the duration of proton flux enhancement is determined by CME propagation timescale $t_{CME}\propto L/V_{CME}$.
\begin{equation}
E_p\propto F_{p}t_{CME}\propto E_{flare}
\end{equation}

From equation (4) and (5) we express $F_{p}$ as follows.
\begin{equation}
F_p\propto E_{flare}^{5/6}\propto V_{CME}^5
\end{equation}
In the derivation, we neglect the proton energy spectral variation depending on flare magnitude.

The scaling relation (6) $F_{p}\propto V_{CME}^5$ is plotted as a dashed line in Figure 2 and compared with the observed correlation. The linear regression fit in double logarithmic space was $F_p \propto V_{CME}^\gamma$ with $\gamma = 4.35\pm 0.50$ which has a slightly smaller slope compared to (6). 

\section{Estimation of CME speed and proton flux associated with solar superflares}
In this section, we compare the scaling relations derived above with observational statistical relations, and try to estimate how fast CME and how intense proton flux will result in the case of solar superflares.

\citet{emslie2012} discussed that kinetic energy of CME is comparable with flare energy, namely $E_{CME} \sim E_{flare}$.

Figure 4 shows the correlation between CME kinetic energy estimated from LASCO observation by $E_{CME}=1/2M_{CME}V_{CME}^2$ and flare SXR peak flux $F_{SXR}$ associated with 80 major SPEs with CME mass estimation.The  correlation coefficient was $r = 0.28$ and the linear regression with OLS bisector method results in $E_{CME} \propto  F_{SXR}^\epsilon$ with $\epsilon = 0.80\pm 0.07$.

In order to estimate the upper-limit of CME speed and energetic proton flux in response to SXR class of solar flares, we try to relate $V_{CME}$ and $F_p$ with $F_{SXR}$ by assuming $F_{SXR}$ is roughly proportional to total released energy during flares, namely $F_{SXR} \propto E_{flare}$.

Based on this assumption, $F_p$ and $V_{CME}$ are respectively scaled with $F_{SXR}$ as
\begin{equation}
V_{CME}\propto F_{SXR}^{1/6},
\end{equation}
\begin{equation}
F_p\propto F_{SXR}^{5/6}.
\end{equation}

Scaling relations (7) and (8) are shown as dashed lines in Figures 3 and 1, respectively. The dashed lines are positioned in each plots so that they pass through the upper-left-most SPE, in order that we can discuss the upper limit of CME speed ($V_{CME, upper limit}$) and proton flux ($F_{p, upper limit}$) in response to $F_{SXR}$. Explicit formulas are as follows, 
\begin{equation}
V_{CME, upper limit} = V_0 F_{SXR}^{1/6},
\end{equation}
\begin{equation}
F_{p, upper limit} = F_{p,0} F_{SXR}^{5/6},
\end{equation}
where $V_0 = 1.3\times 10^4$ km s$^{-1}$, $F_{p,0} = 10^{7.83}$ pfu and $F_{SXR}$ is normalized in unit of 1 W m$^{-2}$. 

Compared with linear regression fits, namely, $V_{CME} \propto F_{SXR}^\alpha$ and $F_p \propto F_{SXR}^\beta$ with $\alpha = 0.30\pm 0.04$, $\beta = 1.19 \pm 0.08$, the derived scaling relations had gentler slopes of $1/6\simeq 0.17$ and $5/6 \simeq 0.83$, respectively. We note that in Figures 3 and 1, the scaling relations (7) and (8) seem consistent with the line of the upper limit of observed $V_{CME}$ and $F_p$ with respect to $F_{SXR}$. 

From equation (9), the upper limit of $V_{CME}$ for X10, X100 and X1000 solar flares will be $V_{CME, X10} = 4.2\times 10^3$ km~s$^{-1}$, $V_{CME, X100} = 6.2\times 10^3$ km~s$^{-1}$ and $V_{CME, X1000} = 9.1\times 10^3$ km~s$^{-1}$, respectively (Figure 5 (a)).  From equation (10), the upper limit of $F_p$ for X10, X100 and X1000 solar flares will be $F_{p, X10} = 2.0\times 10^5$ pfu, $F_{p, X100} = 1.6\times 10^6$ pfu and $F_{p, X1000} = 1.0\times 10^7$ pfu, respectively (Figure 5 (b)).

\section{Impact of superflare-associated CMEs and SPEs on space weather and terrestrial environment}
In this paper we studied CME properties and energetic proton flux associated with possible superflares on the Sun. The scaling relations expressing $M_{CME}$, $V_{CME}$ and $F_p$ in terms of $E_{flare}$ derived from simple assumptions are not inconsistent with statistical relations from solar observation. On the basis of the analysis above, we expect CMEs associated with superflares to be fast and heavy, which will have a huge impact on space weather \citep{AJS1861,JGR2003}.

Huge geomagnetic storms are initiated by magnetic reconnection between injected southward magnetic field of ICMEs $B_{s,ICME}$ and Earth's northward magnetic field \citep{PRL1961,JGR1994}. When ICME magnetic field is northward, no geomagnetic storms occur \citep{GRL1995}.
The magnitude of geomagnetic storms is mainly determined by solar wind westward electric field $E_y\sim V_{ICME,1AU}B_{s,ICME}$, where $V_{ICME,1AU}$ is ICME speed near Earth\citep{bur1975,JGR1994}.
The upperlimit of the magnetic field strength of magnetic cloud $B_{s,MC}$ is estimated by the balance of magnetic pressure and dynamic pressure as $B_{s,MC}^2/8\pi \sim 1/2\rho_{SW} (V_{ICME,1AU}-V_{SW})^2 \sim1/2\rho_{SW} V_{ICME,1AU}^2$, where $\rho_{SW}$ and $V_{SW}$ are density and speed of the solar wind near Earth. If we assume typical value range of solar wind proton number density at 1 AU, namely, $n_p=3-8$ cm$^{-3}$ \citep{schwenn1990}, $B_{s,MC}$ is estimated as $B_{s,MC} \sim (0.08-0.13)~(V_{ICME,1AU}/ 1~kms^{-1})$ nT. This is consistent with observationally known fact that magnetic clouds with higher peak speed ($v_{peak}$) also possess stronger core magnetic field ($B_{peak}$), with observational statistical relation $B_{peak}= 0.047 (v_{peak}/ 1~kms^{-1})$ nT \citep{GRL1998}. The upper limit of westward electric field $E_y$ is estimated as $E_y \sim \sqrt{4\pi\rho_{SW}}V_{CME,1AU}^2$.

CMEs are decelerated during their propagation in the interplanetary space \citep{gopa2001}, sweeping up the interplanetary plasma on their path. We expect from conservation of momentum that if a CME ejecta is heavy enough (comparable to or heavier than the mass scraped up on its path), the CME will not be decelerated much. Fast and heavy ICMEs with southward magnetic field associated with solar superflares would cause extreme geomagnetic storms.

Extreme increase of radiation levels in space associated with solar superflares also result in the increase of radiation at the flight altitude and sometimes at the sea level (Ground Level Enhancement; GLE) through airshower formation in Earth's atmosphere. The radiation levels in Earth atmosphere depend on high energy component of energetic proton flux in space. For example, GLEs are mainly caused by energetic protons injected to the top of the atmospheric layer whose energy is higher than $\sim 1$ GeV. The maximum energy of energetic protons associated with solar flares are known to be less than several GeV.

We estimate the maximum possible energy of energetic protons accelerated at CME-driven shocks. If we apply Hillas limit \citep{hillas1984}, we get the estimation of proton maximum energy as $E_{max}\sim 2$ GeV $B_{0.1}V_{3000}L_{1Rs}$, where $B_{0.1}$, $V_{3000}$ and $L_{1Rs}$ are the upstream magnetic field strength in unit of 0.1 G, shock propagation speed in unit of 3000 km s$^{-1}$ and length scale of acceleration site in unit of the solar radius, respectively. If we assume $L$ is independent of flare energy, we obtain, $E_{max}\propto V_{CME}\propto E_{flare}^{1/6}$. On the other hand, if we assume protons with highest energies are accelerated by the electric voltage generated by magnetic reconnection at flare site, we obtain $E_{max} \sim 7$ GeV $B_{100}V_{100}L_{0.1Rs}$, where $B_{100}$, $V_{100}$ and $L_{0.1Rs}$ are the active region magnetic field strength in unit of 100 G, plasma 'inflow' speed in unit of 100 km s$^{-1}$ and length scale of the flaring active region in unit of 0.1 solar radius, respectively.
Generally, magnetic reconnection rate $\sim V_{in}/V_{CME}$ is independent of flare magnitude \citep{shibm2011}. Applying equation (2) and (4), we obtain $E_{max}\propto E_{flare}^{1/2}$. X1000 class flare, for example, may produce 10 GeV protons which result in drastic increase of radiation level in Earth atmospheric layer.

In the derivation of scaling relation (6), we assumed the relation $t_{CME}\propto L/V_{CME}$, where L is the active region size. If we use constant length scale $L_0$ in place of $L$, the scaling relation would change to $F_p\propto E_{flare}^{7/6}\propto V_{CME}^7$. This might be the case if the particles are accelerated at the shock front in the interplanetary space where physical quantities do not depend on active region size.

\acknowledgements
The CME catalog we used in this study is generated and maintained at the CDAW Data Center by NASA and The Catholic University of America in cooperation with the Naval Research Laboratory. SOHO is a project of international cooperation between ESA and NASA. We studied SPEs from online SPE catalog provided by the CDAW Data Center\footnote{http://cdaw.gsfc.nasa.gov/CME\_list/sepe/}. This work was supported by JSPS KAKENHI Grant Numbers 16H03955. The authors are grateful to Dr. Seiji Yashiro for providing valuable information and giving us fruitful comments on data handling. This work is motivated partly by Mr. Taira Hiraishi's master thesis.



\begin{figure}
\epsscale{.90}
\plotone{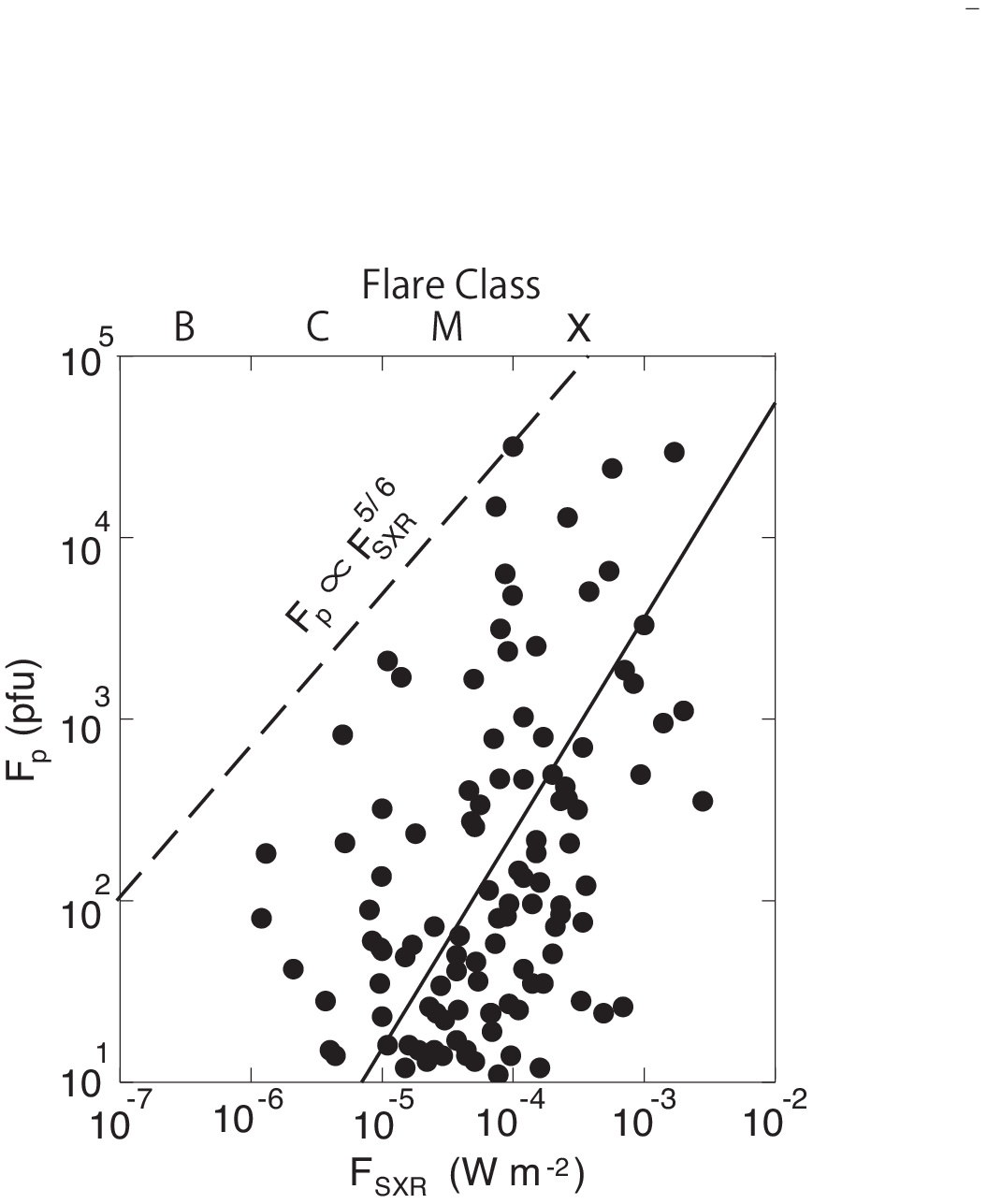}
\caption{$F_p$-$F_{SXR}$ relation. Solid line is the linear regression fit, of equation $F_p \propto F_{SXR}^\beta$ with $\beta = 1.19\pm 0.08$. The dashed line is the upper limit of $F_p$ in terms of $F_{SXR}$ (equation (10)).
\label{flare}}
\end{figure}

\begin{figure}
\epsscale{.90}
\plotone{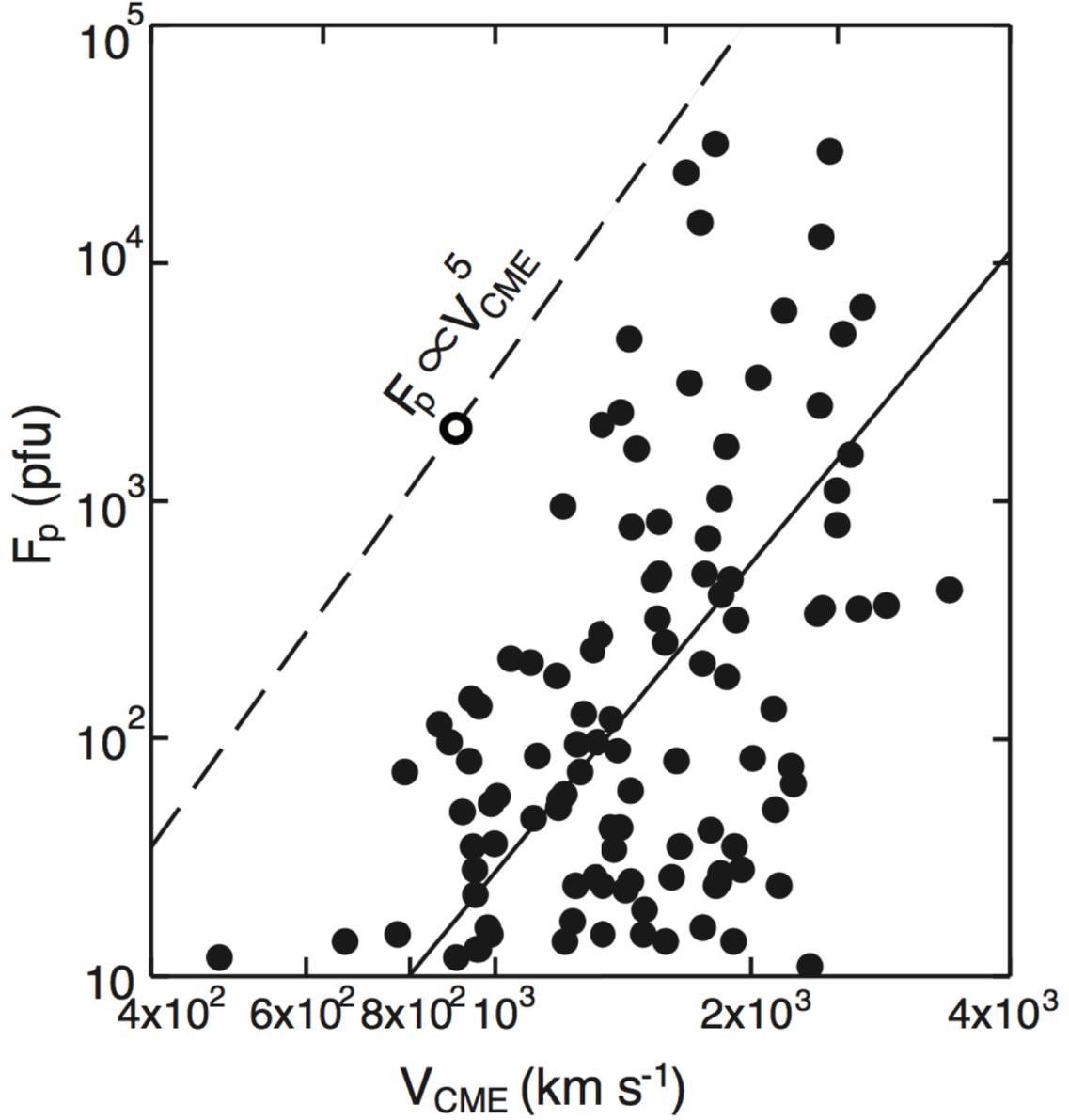}
\caption{$F_p$-$V_{CME}$ relation. Solid line is the linear regression fit $F_p \propto V_{CME}^\gamma$ with $\gamma = 4.35\pm 0.50$. The dashed line is the upper limit of $F_p$ in terms of $V_{CME}$ whose spectral index is 5. The SPE represented by an unfilled circle in the figure seems to be an outlier, and the linear regression without it made the slope a bit steeper, namely $\gamma =4.50\pm 0.48$.
\label{vp}}
\end{figure}

\begin{figure}
\epsscale{.90}
\plotone{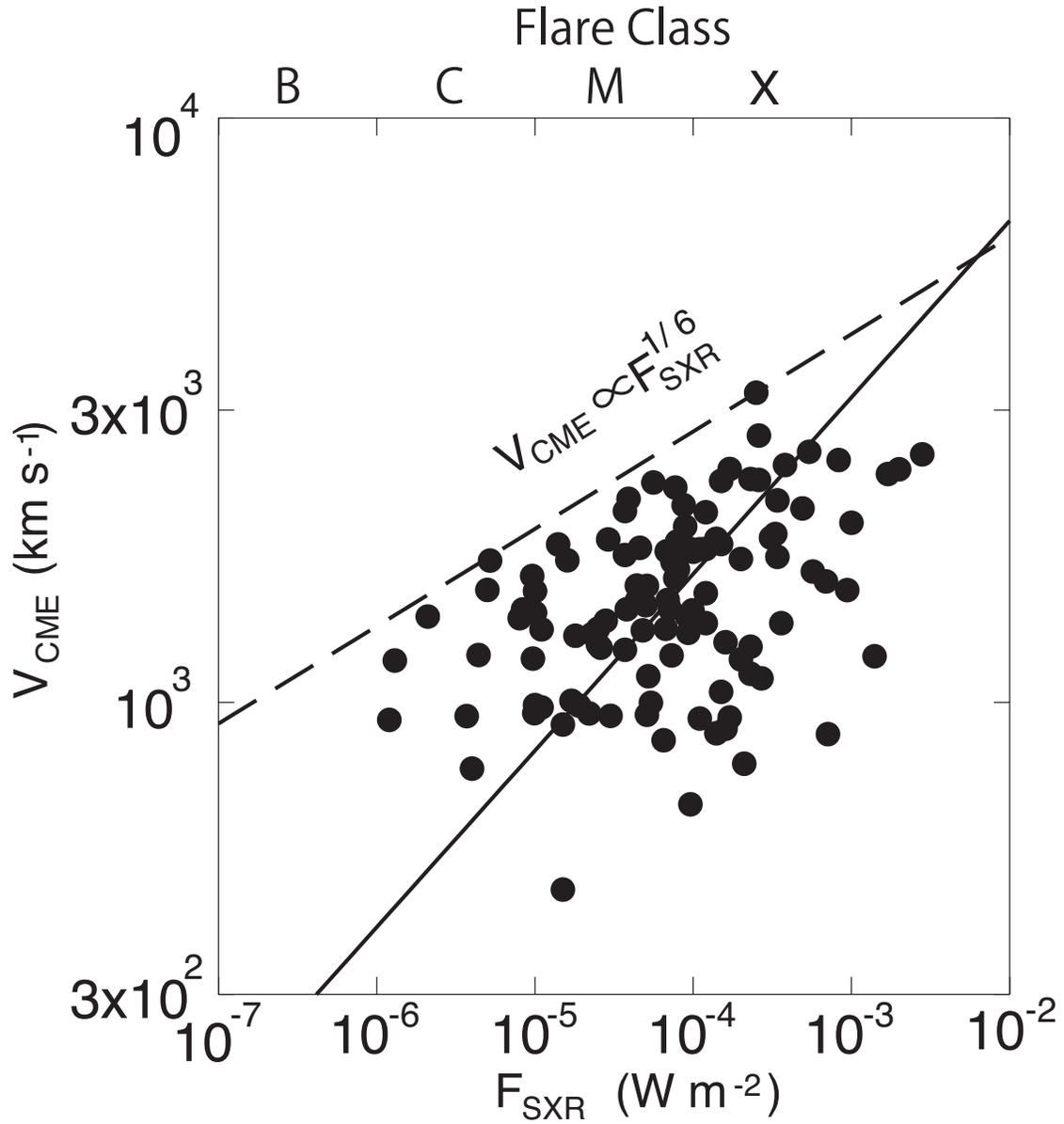}
\caption{$V_{CME}$-$F_{SXR}$ relation. Solid line is the linear regression fit, of equation $V_{CME} = F_{SXR}^\alpha$ with $\alpha = 0.30\pm 0.04$. The dashed line is the upper limit of $V_{CME}$ in terms of $F_{SXR}$ (equation (7)).
\label{flare}}
\end{figure}

\begin{figure}
\epsscale{.90}
\plotone{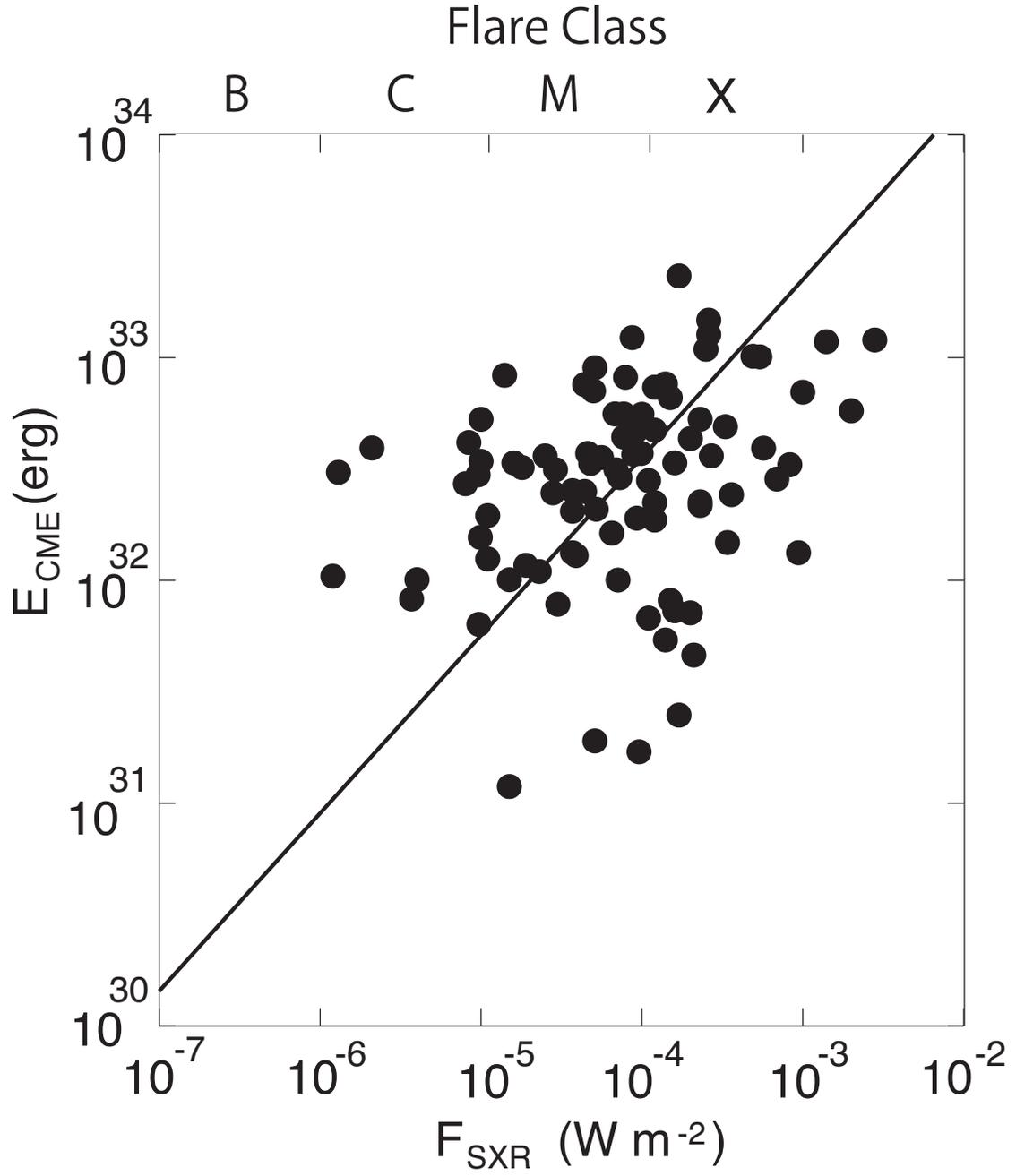}
\caption{$E_{CME}$-$F_{SXR}$ relation. Solid line is the linear regression fit, of equation $E_{CME} \propto F_{SXR}^\epsilon$ with $\epsilon = 0.80\pm 0.07$.
\label{flare}}
\end{figure}

\begin{figure}
\epsscale{.90}
\plotone{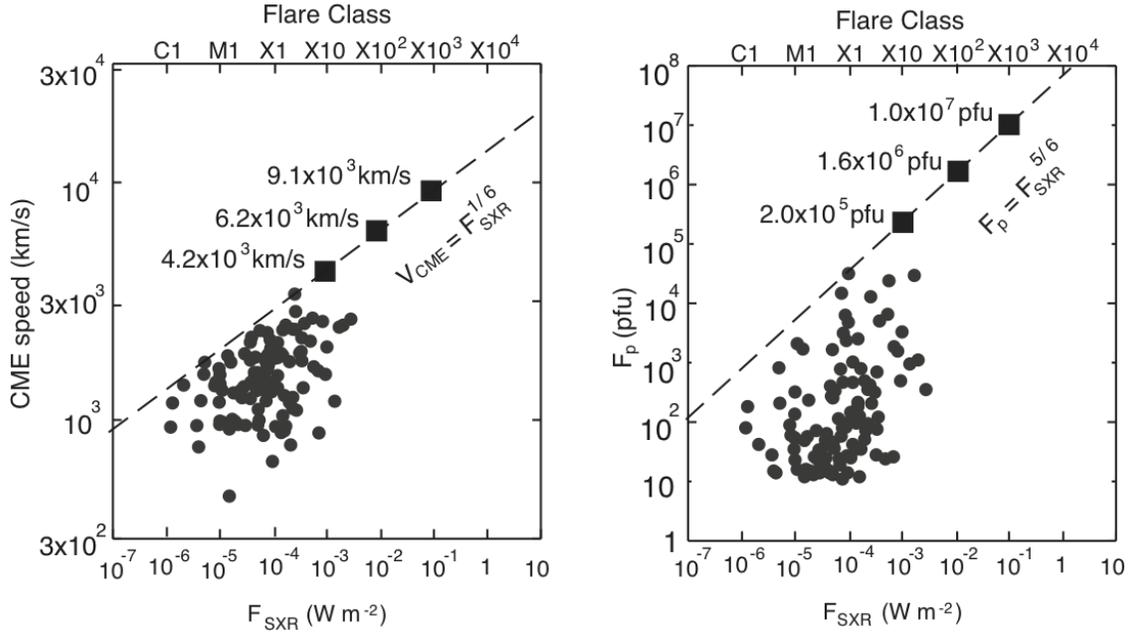}
\caption{(a): $V_{CME}$-$F_{SXR}$ relation. The estimated upperlimits of CME speed associated with X10, X100 and X1000 class flares are represented as black filled rectangles. (b): $F_p$-$F_{SXR}$ relation. The estimated upperlimits of energetic proton flux associated with X10, X100 and X1000 class flares are represented as black filled rectangles. The dashed lines in (a) and (b) are the upperlimits of $V_{CME}$ and $F_p$ in terms of $F_{SXR}$ given by equation (4) and (6), respectively.
\label{flare}}
\end{figure}


\begin{thebibliography}{}
\bibitem[Aarnio et al.(2012)]{aar2012}
Aarnio, A.~N., Matt, S.~P., \& Stassun, K.~G.\ 2012, \apj, 760, 9 
%
\bibitem[Aarnio et al.(2011)]{aar2011}
Aarnio, A.~N., Stassun, K.~G., Hughes, W.~J., \& McGregor, S.~L.\ 2011, \solphys, 268, 195 
%
\bibitem[Airapetian et al.(2016)]{air2016}
Airapetian, V. S., Glocer, A., Gronoff, G. et al.\ 2016, Nature Geoscience, 9, 452
%
\bibitem[Aschwanden(2012)]{asch2012}
Aschwanden, M.~J.\ 2012, \ssr, 171, 3 
%
%
%
\bibitem[Burton et al.(1975)]{bur1975}
Burton, R., R. McPherron, and C. Russell \ 1975, Journal of Geophysical Research (Space Physics), 80, 4204-4214
%
%
\bibitem[Dungey(1961)]{PRL1961}
Dungey, J.~W.\ 1961, Physical Review Letters, 6, 47 
%
%
\bibitem[Emslie et al.(2012)]{emslie2012}
Emslie, A.~G., Dennis, B.~R., Shih, A.~Y., et al.\ 2012, \apj, 759, 71 
%
\bibitem[Gonzalez et al.(1994)]{JGR1994}
Gonzalez, W.~D., Joselyn, J.~A., Kamide, Y., et al.\ 1994, \jgr, 99, 5771 
%
\bibitem[Gonzalez et al.(1998)]{GRL1998}
Gonzalez, W.~D., de Gonzalez, A.~L.~C., Dal Lago, A., et al.\ 1998, \grl, 25, 963 

\bibitem[Gopalswamy et al.(2001)]{gopa2001}
Gopalswamy, N., Lara, A., Yashiro, S., Kaiser, M.~L., \& Howard, R.~A.\ 2001, \jgr, 106, 29207
%
\bibitem[Gopalswamy et al.(2003)]{gopa2003a}
Gopalswamy, N., Yashiro, S., Kaiser, M.~L., \& Howard, R.~A.\ 2003, Advances in Space Research, 32, 2613 

\bibitem[Gopalswamy et al.(2003)]{gopa2003b}
Gopalswamy, N., Yashiro, S., Lara, A., et al.\ 2003, \grl, 30, 8015 
%
\bibitem[Gopalswamy et al.(2004)]{gopa2004}
Gopalswamy, N., Yashiro, S., Krucker, S., Stenborg, G., \& Howard, R.~A.\ 2004, Journal of Geophysical Research (Space Physics), 109, A12105 
%
\bibitem[Gopalswamy(2009)]{gopa2009}
Gopalswamy, N.\ 2009, Climate and Weather of the Sun-Earth System (CAWSES)

\bibitem[Gopalswamy et al.(2010)]{gopa2010a}
Gopalswamy, N., Akiyama, S., Yashiro, S., \& M{\"a}kel{\"a}, P.\ 2010, Astrophysics and Space Science Proceedings, 19, 289 


\bibitem[Gopalswamy(2011)]{gopa2011}
Gopalswamy, N.\ 2011, Astronomical Society of India Conference Series, 2,  

\bibitem[Gopalswamy et al.(2012)]{gopa2012}
Gopalswamy, N., Xie, H., Yashiro, S., et al.\ 2012, \ssr, 171, 23 
%
\bibitem[Gopalswamy et al.(2014)]{gopa2014}
Gopalswamy, N., Xie, H., Akiyama, S., M{\"a}kel{\"a}, P.~A., \& Yashiro, S.\ 2014, Earth, Planets, and Space, 66, 104 
%
%
\bibitem[Gopalswamy et al.(2015)]{gopa2015c}
Gopalswamy, N., Yashiro, S., Xie, H., Akiyama, S., \& M{\"a}kel{\"a}, P.\ 2015, Journal of Geophysical Research (Space Physics), 120, 9221 
%
%
\bibitem[Hillas(1984)]{hillas1984}
Hillas, A.~M.\ 1984, \araa, 22, 425
%
\bibitem[Howard et al.(1982)]{how1982}
Howard, R.~A., Michels, D.~J., Sheeley, N.~R., Jr., \& Koomen, M.~J.\ 1982, \apjl, 263, L101 
%
\bibitem[Hudson(2011)]{hud2011}
Hudson, H.~S.\ 2011, \ssr, 158, 5 
%
\bibitem[Illing \& Hundhausen(1986)]{JGR1986}
Illing, R.~M.~E., \& Hundhausen, A.~J.\ 1986, \jgr, 91, 10951 
%
\bibitem[Isobe et al.(1990)]{iso1990}
Isobe, T., Feigelson, E.~D., Akritas, M.~G., \& Babu, G.~J.\ 1990, \apj, 364, 104 
%
\bibitem[Kennel et al.(1984a)]{JGR1984a}
Kennel, C.~F., Edmiston, J.~P., Russell, C.~T., et al.\ 1984, \jgr, 89, 5436
\bibitem[Klein \& Burlaga(1982)]{JGR1982}
Klein, L.~W., \& Burlaga, L.~F.\ 1982, \jgr, 87, 613
%
\bibitem[Kurt et al.(2004)]{kurt2004}
Kurt, V., Belov, A., Mavromichalaki, H., \& Gerontidou, M.\ 2004, Annales Geophysicae, 22, 2255 
%
\bibitem[Livingston et al.(2006)]{liv2006}
Livingston, W., Harvey, J.~W., Malanushenko, O.~V., \& Webster, L.\ 2006, \solphys, 239, 41 
%
\bibitem[Loomis(1861)]{AJS1861}
Loomis, Elias, \ 1861, American Journal of Science 96, 318
%
\bibitem[Maehara et al.(2012)]{mae2012}
Maehara, H., Shibayama, T., Notsu, S., et al.\ 2012, \nat, 485, 478 
%
\bibitem[Miyake et al.(2012)]{miya2012}
Miyake, F., Nagaya, K., Masuda, K., \& Nakamura, T.\ 2012, \nat, 486, 240 
%
\bibitem[Neupert(1968)]{neu1968}
Neupert, W. M. 1968, \apj, 153, L59
%
\bibitem[Reagan et al.(1973)]{reagan1973}
Reagan, Joseph B., William L. Imhof, \& Vincent F. Moughan. 1973, LOCKHEED MISSILES AND SPACE CO INC PALO ALTO CA PALO ALTO RESEARCH LAB, No. LMSC-D352207
%
\bibitem[Reames et al.(1996)]{reames1996}
Reames, D.~V., Barbier, L.~M., \& Ng, C.~K.\ 1996, \apj, 466, 473 
%
\bibitem[Reames(2000)]{reams2000}
Reames, D.~V.\ 2000, 26th International Cosmic Ray Conference, ICRC XXVI, 516, 289 
%
%
%
\bibitem[Schrijver et al.(2012)]{schrij2012}
Schrijver, C.~J., Beer, J., Baltensperger, U., et al.\ 2012, Journal of Geophysical Research (Space Physics), 117, A08103 
%
\bibitem[Schwenn(1990)]{schwenn1990}
Schwenn, R.\ 1990, Physics of the Inner Heliosphere I, 99
%
\bibitem[Shibata \& Magara(2011)]{shibm2011}
Shibata, K., \& Magara, T.\ 2011, Living Reviews in Solar Physics, 8, 6 

\bibitem[Thomson et al.(2004)]{GRL2004}
Thomson, N.~R., Rodger, C.~J., \& Dowden, R.~L.\ 2004, \grl, 31, L06803
%
\bibitem[Tsurutani et al.(1988)]{JGR1988}
Tsurutani, B.~T., Smith, E.~J., Gonzalez, W.~D., Tang, F., \& Akasofu, S.~I.\ 1988, \jgr, 93, 8519 
%
\bibitem[Tsurutani \& Gonzalez(1990)]{JGR1990}
Tsurutani, B.~T., \& Gonzalez, W.~D.\ 1990, \jgr, 95, 12305
%
\bibitem[Tsurutani \& Gonzalez(1995)]{GRL1995}
Tsurutani, B.~T., \& Gonzalez, W.~D.\ 1995, \grl, 22, 663 
%
\bibitem[Tsurutani \& Lin(1985)]{JGR1985}
Tsurutani, B.~T., \& Lin, R.~P.\ 1985, \jgr, 90, 1 
%
\bibitem[Tsurutani et al.(2003)]{JGR2003}
Tsurutani, B.~T., Gonzalez, W.~D., Lakhina, G.~S., \& Alex, S.\ 2003, Journal of Geophysical Research (Space Physics), 108, 1268 
\bibitem[Tsurutani et al.(2009)]{RS2009}
Tsurutani, B.~T., Verkhoglyadova, O.~P., Mannucci, A.~J., et al.\ 2009, Radio Science, 44, RS0A17 
%
%
%
%
%
\bibitem[Yashiro et al.(2004)]{yas2004}
Yashiro, S., Gopalswamy, N., Michalek, G., et al.\ 2004, Journal of Geophysical Research (Space Physics), 109, A07105 
%
\bibitem[Yashiro et al.(2006)]{yas2006}
Yashiro, S., Akiyama, S., Gopalswamy, N., \& Howard, R.~A.\ 2006, \apjl, 650, L143 
%
\bibitem[Yashiro \& Gopalswamy(2009)]{yas2009}
Yashiro, S., \& Gopalswamy, N.\ 2009, Universal Heliophysical Processes, 257, 233 
%
\bibitem[Zhang et al.(2007)]{zhang2007}
Zhang, J., Richardson, I.~G., Webb, D.~F., et al.\ 2007, Journal of Geophysical Research (Space Physics), 112, A10102

\end{thebibliography}
\end{document}